\documentclass{amsart}%
\usepackage{amsfonts}
\usepackage{amsmath}
\usepackage{amssymb}
\usepackage{graphicx}%
\theoremstyle{plain}

\numberwithin{equation}{section}
\begin{document}
\title[ Derivation of index theorem by supersymmetry]{Derivation of index theorem by supersymmetry}
\author[Da Xu]{Da Xu}
\address{Department of Mathematics\\
The University of Iowa\\
Iowa City, IA 52242} \email{dxu@math.uiowa.edu}
\keywords{supersymmetry, index theorem, nonlinear $\sigma$-model}
\maketitle

 \begin{abstract}
   The present paper gives calculations in detail to prove several special
   cases of Atiyah-Singer theorem through supersymmetric $\sigma$-models.
   Some technical tricks are employed to calculate the
   determinants of fluctuation operators of the path integrals. An
   intuitive and geometric argument is applied to overcome the
   complicated calculation on spin fields twisted by gauge fields.
\end{abstract}

 \maketitle
\section{introduction}

      Topological invariants were defined in \cite{W1} for the
     theories in which supersymmetry is not spontaneously broken.It
     was suggested in that paper that the topological invariant
     $Tr (-1)^F$ plays an important role in the index theorem, which is
     an modern development of Riemann-Roch theorem. Since then,
     there had been a lot of intensive work on the relationship
     between supersymmetry and topological information of manifolds. Another influential paper
     \cite{W2} showed that Morse theory can be derived as conclusions of
     a supersymmetric theory.

        On the other hand, it had been well known that anomalies in many situations are
    intimately connected to the index theorem. Fujikawa gave a
    very direct method to derive the gauge anomaly of a spin field,
    which explains the topological origin of anomalies  most
    clearly \cite{F1},\cite{F2}.Gratitational anomaly was derived in \cite{AG} using Fujikawa's
    method. \cite{AW} pointed out the relationship between gravitational
    anomalies and the index theorem. Gravitational anomalies of spin $\frac{1}{2}$ and spin
    $\frac{3}{2}$ fermions and self-dual gauge fields are considered in \cite{AW}.  Then several papers were also
    concerned with this topic.\cite{AG} gave the method to calculate the
    index density for different complexes. Although the idea is simple, the calculations to realize the idea are complicated.
    \cite{N} gave the calculation of Dirac genus by considering a
    $0+\frac{1}{2}$ nonlinear $\sigma$ model. \cite {M1},\cite{M2} gave the same
    result by using Green function which is essentially the same as the method of heat kernal.
       The present paper derives the Hirzebruch signature theorem
       and the Hirzebruch-Riemann-Roch theorem in detail. In the
       derivation of the Hirzebruch-Riemann-Roch theorem, we need to
       consider a supersymmetric theory coupled with gauge field.
       Some papers got the index of twisted spin field (twisted by
       gauge field) using very complicated calculation. However we
       will give a very simple idea, which can be applied to any gauge coupled index, to overcome this difficulty.In the
   first section of the present paper, we review the procedure of
   calculating index by supersymmetric theories. In the second
   section, we give the full calculation of the Hirzebruch signature
   theorem and Hirzebruch-Riemann-Roch theorem. In the last section
   the conclusion is drawn and further questions are asked.

\section{the procedure of calculating indice}
 In this section, let's review the basic procedure to calculate the
 index by a nonlinear supersymmetric $\sigma$ model. We will see how this model is connected the
 Atiyah-Singer theorem. We follow the
 literatures on this topic which are mentioned in the introduction
 part of this paper.
    Suppose we have supersymmetric theory and the supercharges are
 $Q_1,Q_2,...,Q_k$ satisfying
 \begin{align}
  & Q_1^2=Q_2^2=...=Q_k^2=H,\nonumber\\
  &  Q_iQ_j+Q_jQ_i=0 (i\neq j).
 \end{align}
 We can choose one $Q_i$ and denote it by $Q$. An important
 fact is that $Q$ pairs a nonzero energy bosonic state with a
 fermionic state. Specifically, let $|b>$ be a bosonic state with a
 nonzero energy $E$. Then define a fermionic state $|f>=1/\sqrt{E}$.
 Since $Q^2=H$, $|f>$ also satisfies $Q |f>=\sqrt{E} |b>$ and
 $H |f>=E |f>$.
    On the other hand,obviously, if $E$ is zero, the construction
  above can't hold which means zero energy states can't be paired
   as bosons and fermions. We denote the number of zero-energy bosonic states by
   $n_B^0$ and the number of zero-energy fermionic states by $n_F^0$. As
  the parameters of the theory vary,  a nonzero state maybe becomes a zero state.
  Its superpartner must also become a zero state. In the case, both $n_B^0$ and $n_F^0$ increase by 1. On the other hand,
  if a zero-energy state becomes a non-zero energy state, since
  every non-zero state must be paired with its superpartner, there
  must also exist another zero-energy state,which is the
  superpartner of the zero-energy state mentioned, becoming a
  non-zero energy state. In one word, $n_B^0-n_F^0$ is an invariant
  of the parameters of the theory. We introduce the operator
  $(-1)^F$, where $F$ is the fermionic number operator. Then it is
  obvious that
  \begin{equation}
  n_B^0-n_F^0=Tr (-1)^F.
  \end{equation}
  In addition, the following important equality holds because
  non-zero energy states are paired.
  \begin{equation}
  Tr (-1)^F e^{-\beta H},
  \end{equation}
  where $\beta \leq 0$.
  The above doesn't depend on $\beta$.

     The Hilbert space of the theory can be split into the bosonic
     subspace $H_B$ and the fermionic subspace $H_F$. Therefore $Q$
     can be written in the form
    \begin{equation}
    Q=\left(
       \begin{array}{cc}
         0 & D^+ \\
         D & 0 \end{array}\right ).
     \end{equation}
 Now we can see that $n_B^0$ is the dimension of the space
 ${D\phi=0}$ and the $n_F^0$ is the dimension of the space
 ${D^+ \phi=0}$.

 By above, we have
 \begin{equation}
 ind D=\ker D-\ker D^+=Tr (-1)^F e^{-\beta H}
 \end{equation}
 The right hand side of (2.5) can be calculated by path integrals if
 we change the time to negative imaginary time. Since the existence
 of $(-1)^F$ in the right hand side of (2.5), we need to impose
 periodic conditions on bosonic varialbes and fermionic
 variables.\newpage

 Before we get into next section, let's
 briefly go over the Atiyah-Singer theorem \cite{AS1},\cite{AS2},\cite{ASe}:

 Let $(E,D)$ be an elliptic complex over an m-dimensional compact
manifold M without a boundary.The index of this complex is given by
$$ ind(E,D)=(-1)^{m(m+1)/2}\int_M
ch(\oplus_r (-1)^r E_r)\frac{Td(TM^\mathbb{C})}{e(TM)}|_{vol}.
$$
This theorem unifies all the underlying theorems. There have been
different mathematical proofs for this theorem based on $K$ theory,
heat kernal and etc.

\section{calculations on different complexes}

\subsection{Hirzebruch signature theorem}

   Let us review the formula of Hirzebruch signature.
   Let $M$ be a
   compact orientable manifold of even dimension which is
   $m=2l$.Then we can define a bilinear map
   $H^l(M;R)\times H^l(M;R)\rightarrow R$ by\\
   $$
   \sigma([\omega],[\eta])=\int_M \omega\wedge\eta.
   $$
   By Poincar$\acute{e}$ duality,$\sigma$ is actually a non-degenerate.
   If $l$ is an even number, $sigma$ is symmetric obviously.Thus it
   doesn't have zero eigenvalue. Therefore the sum of the number of
   positive eigenvalues $b^+$ and the number of negative eigenvalues
   $b^-$ is $l$. The Hirzebruch signature is defined by
   $$
   \tau(M)\equiv b^{+}-b^-.
   $$
   If $l$ is odd, $\tau(M)$ is defined to be zero.Now consider the
   operator which is
   $$
   \mathfrak{D}=d+d^{+}.
   $$
   We also define an operator
   $\pi:\Omega^r(M)^\mathbb{C}\rightarrow\Omega^{m-r}(M)^\mathbb{C}$by
   $$
   \pi\equiv i^{r(r-1)+l}*.
   $$
   It is easy to check that  $\pi^2=1$ and that $\pi$ anticommutes with $\mathfrak{D}$. It has two eigenvalues which are 1 and -1.
   Now the space can be written in the form of a direct sum of the
   two eigenspaces:
   $$
   \Omega^*(M)^\mathbb{C}=\Omega^+(M)\oplus\Omega^-(M).
   $$
   Now the signature complex is defined by the restriction of
   $\mathfrak{D}$ on $\Omega^+(M)$ which is
   $$
   \mathfrak{D}_+:\Omega^+(M)\rightarrow\Omega^-(M).
   $$
   If we apply Atiya-Singer index theorem to the signature
   complex, we can get the index of $\mathfrak{D}_+$ immediately:
   \begin{align}
   ind  \mathfrak{D}_+&=(-l)^l \int_M ch(\wedge^+
   T^*M^\mathbb{C}-\wedge^-TM^*\mathbb{C})\frac{Td(TM^\mathbb{C})}{e(TM)}\mid_{vol}\nonumber\\
   &=2^l \int_M \prod_{i=l}^l \frac{x_i/2}{\tanh
   x_i/2}|_{vol}\nonumber\\
   &=\int_M \prod_{i=1}^l \frac{x_i}{\tanh x_i}|_{vol}\nonumber\\
   &=\int_M L(TM)|_{vol}.
   \end{align}
   The last equality is the definition of L-polinomial. It is
   easy to see that if $m\equiv 2 \mod 4$, $\tau(M)$ is zero.

   Now let us see how we can derive this formula for a $0+1$
   nonlinear $\sigma$ model. consider a theory which is
   \begin{equation}
   L=\frac{1}{2}g_{ij}(x)\dot{x}^i
   \dot{x}^j+\frac{1}{2}g_{ij}(x)\bar{\psi}^i\gamma^0\frac{D}{dt}\psi^j
   +\frac{1}{12}R_{ijkl}\bar{\psi}^i\psi^k\bar{\psi}^j \psi^l.
   \end{equation}
   The hamiltonian for this lagrangian is
   \begin{equation}
   H=\frac{1}{2}g_{ij}(x)\dot{x}^i
   \dot{x}^j-\frac{1}{4}R_{ijkl}\bar{\psi}^i\psi^k\bar{\psi}^j
   \psi^l.
   \end{equation}
   The lagrangian has a discrete symmetry:
   $$
   \psi^i\rightarrow \gamma_5\psi^i.
   $$
   As pointed out in \cite{AG}, $Tr Q_5 e^{-\beta H}$ depends on only zero
   modes and is independent of $\beta$.
   The supersymmetric transformations are
   \begin{align}
   \delta x^i&=\bar\epsilon^i,\nonumber\\
   \delta \psi^i&=\gamma^0\dot{x}^i
   \epsilon-\Gamma^i_{ik}\bar\epsilon \psi^j \psi^k.
   \end{align}

    $Tr Q_5 e^{-\beta H}$ can be written in terms of path integral
    \begin{equation}
    Tr Q_5 e^{-\beta H}=\int_{PBC}Dx D\psi_2 \int_{ABC} D\psi_1
    e^{(-S_E)},
    \end{equation}
    where $\psi_1$ and $\psi_2$ belong to eigenspaces of eigenvalues
    1 and -1 of $\gamma_5$.
     Notice we have to impose antiperiodic boundary condition on
     fermionic variables when we calculate tr. Therefore we have the
     above boundary condition because  $Q_5$ is included in the
     calculation.
     Since $x^i$s and $\psi_2^i$s are periodic functions, we can
     expand them by fourier series
     $$
     x^i(t)=\frac{1}{\sqrt\beta}\Sigma_{n=-\infty}^{\infty}\xi_n^i
     e^{\frac{2\pi int}{\beta}},\\
     \psi_2^i=\frac{1}{\sqrt\beta}\Sigma_{n=-\infty}^{\infty}\psi_n^i
     e^{\frac{2\pi int}{\beta}}.
     $$
     We expand the action $S$ around constant $(x,\psi_2)$
     (fix$\psi_1$)
     \begin{align}
     L^{(2)}=&\frac{1}{2}g_{ij}(x_0)\dot{x}^i \dot{x}^j +\frac{1}{2}(\psi_2^i-\psi_{20}^i)\frac{d}{dt}(\psi_2^j-\psi_{20}^j)g_{ij}(x)
     +\frac{1}{2}\psi_1^i \frac{d}{dt}\psi_1^j g_{ij}(x_0)\nonumber\\
     &+\frac{1}{4}R_{ijkl}\psi_{20}^k\psi_{20}^l \psi_1^i
     \psi_1^j
     +\frac{1}{4}R_{ijkl}\psi_{20}^k\psi_{20}^l
     (x^i-x^i_0)(x^j-x^j_0)\psi_1^j;
     \end{align}
     the fluctuation operator for $\delta x=x-x_0$ is
     \begin{equation}
     -\delta_{ij}\frac{d^2}{dt^2}+\frac{1}{2}R_{ijkl}\psi_{20}^k\psi_{20}^l\frac{d}{dt}.
     \end{equation}
     The fluctuation operator for $\delta\psi_2=\psi_2-\psi_{20}$ is
     \begin{equation}
     \delta_{ij}\frac{d}{dt};
     \end{equation}
     the fluctuation operator for $\psi_1$ is (we need to remind
     that the boundary condition for this operator is antiperiodic)
     \begin{equation}
     \frac{d}{dt}+\frac{1}{2}R_{ijkl}\psi_{20}^k \psi_{20}^l.
     \end{equation}
     Now we can calculate the index of the operator $\mathbb{D}_+$.
     We have
     \begin{align}
     Ind \mathbb{D}_+&=\mathcal{N}\int\prod_{\mu=1}^d
     \frac{x_0}{\sqrt{2\pi}}d \psi_{20} d\psi_1^\mu
     [Det_{PBC'}(-\delta_{ij}\frac{d^2}{dt^2}+\frac{1}{2}\mathfrak{R}_{ij}\frac{d}{dt})]^{-\frac{1}{2}}\times\nonumber\\
     [Det_{PBC'}(\delta_{ij}\frac{d}{dt})]^{\frac{1}{2}}\times
     &[Det_{APBC'}(\frac{d}{dt}+\frac{1}{2}\mathfrak{R}_{ij})^{\frac{1}{2}}\nonumber\\
     &=\mathcal{N}\int\prod_{\mu=1}^d
     \frac{x_0}{\sqrt{2\pi}}d \psi_{20} d\psi_1^\mu.
     \end{align}
     In order to calculate the right hand side of the equality
     above, we need to calculate $\mathfrak{N}$ and the three
     determinants.
     First, let us calculate $\mathfrak{N}_b$ which is the
     normalization factor of the second determinant in the
     integral. By the amplitude $<x,0|x,1>=\frac{1}{\sqrt 2\pi}$, we
     have a path integral
     \begin{align}
     \int\mathcal{N}x_\mu e^{-\frac{1}{2}\int_0^1 \dot{x}^{\mu 2}}
     &=\mathcal{N}(Det_{\delta_{\mu\nu}+\partial_t^2})
     ^{-\frac{1}{2}}
      \int\Pi_\mu^{2n}\frac{dx^\mu}{\sqrt{2\pi}}\nonumber\\
     &=(\sqrt{2\pi})^{-n}\int\prod_\mu^{2n}dx^\mu.
     \end{align}
     Since the eigenvalues of $-\frac{d^2}{dt^2}$ are
     $\lambda_n=(\frac{2n\pi}{\beta})^2$, we have
     \begin{equation}
     Det'_{PBC}(-\frac{d^2}{dt^2})=\prod_{n\in\mathbb{Z},n\neq
     0}(\frac{2\pi n}{\beta})^2.
     \end{equation}
     The spectral $\zeta$-function is

     \begin{equation}
     \zeta_{-d^2/dt^2}(s)=\sum_{n\in\mathbb{Z},n\neq 0}^\infty[(\frac{2n\pi}{\beta})^2]^{-s}
     =2(\frac{\beta}{2\pi})^{2s}\zeta(2s).
     \end{equation}
     We have
     \begin{align}
     \zeta'_{-d^2/dt^2}(0)&=4\log (\beta/2\pi)e^{2s
     \log(\beta/2\pi)}\zeta(2s)+4e^{2s\log(\beta/2\pi)}\zeta'(2s)|_{s=0}\nonumber\\
     &=4[\log (\beta/2\pi)\zeta(0)+\zeta'(0)]=-2\log\beta.
     \end{align}
     Hence the determinant is
     \begin{equation}
     Det'_{PBC}(-\frac{d^2}{dt^2})=e^{-\zeta'_{-d^2/dt^2}(0)}=\beta^2.
     \end{equation}
     If $\beta=1$, $Det'_{PBC}(-d^2/dt^2)=1$.Thus $\mathcal{N}_b=1$.
     Let us evaluate $\mathcal{N}_{\psi_2}$.
     We know that as a fluctuation operator, $Det'_{PBC}(\delta_{ij})>0$.
     Therefore
     \begin{equation}
     Det'_{PBC}(\delta_{ij}d/dt)=|Det'_{PBC}(\delta_{ij})d^2/dt^2|^{\frac{1}{2}}
     =1.
     \end{equation}
     By the theorem, we know that
     \begin{align}
     Tr \gamma_{2n+1}&=\int_{PBC}\mathcal{D}\psi
     e^{-\frac{1}{2}\int_0^1 \psi\dot{\psi}dt}\nonumber\\
     &=\mathcal{N}_{\psi_2}Det'_{PBC}(\delta_{ij}d/dt)^{\frac{1}{2}}\int
     d\psi_{02}^1...d\psi_{02}^{2n}\nonumber\\
     &=\mathcal{N}\int d\psi_{02}^1...d\psi_{02}^{2n}.
     \end{align}
     We know that
     \begin{equation}
     \gamma_{2n+1}=i^n \gamma_0^1 ...\gamma_0^{2n}=(2i)^n
     \psi_{02}^1...\psi_{02}^{2n},
     \end{equation}
     and $Tr \gamma_{2n+1}^2=Tr I=2^n$.
     Then we have the equation
     \begin{align}
     2^n &=Tr \gamma_{2n+1}^2=\mathcal{N}_{\psi_2}\int d\psi_{02}^1
     ...d\psi_{02}^{2n} (2i)^n
     \psi_{02}^1...\psi_{02}^{2n}\nonumber\\
     &=(2i)^n \mathcal{N}_{\psi_2}.
     \end{align}
     Thus $\mathcal{N}_{\psi_2}=i^n$.

     Let us evaluate the third normalization factor
     $\mathcal{N}_{\psi_1}$.

     We only need to evaluate $Det_{APBC}(d/dt)$.Then immediately we
     can use the same method above.
     We add a harmonic oscillator $\omega$ to the operator. As in \cite{N}, the partition function is
     \begin{equation}
     Tr (-1)^F e^{-\beta
     H}=2\cosh(\beta\omega/2)=e^{\beta\omega/2}Det'_{APBC}((1-\epsilon\omega)d/dt+\omega).
     \end{equation}
     Therefore, as $\omega\rightarrow 0$,
     \begin{equation}
     Det'_{APBC}(d/dt)=\lim_{\omega\rightarrow
     0}e^{-\beta\omega/2}2\cosh(\beta\omega/2)=1.
     \end{equation}

     After evaluating the normalization factor, we need to evaluate
     the determinant in ..
     $\mathfrak{R}$ can be considered as a curvature two form which
     is a antisymmetric tensor, because of $\psi_{02}^k$s are fermionic variables and
     anticommute. We can choose some local coordinates such that
     $\mathfrak{R}$ is diagonal of antisymmetric 2 by 2 matices with vanishing diagonal entries.
     For the $i$th 2$\times$2 matrix, the determinant is

    \begin{align}
      I(\beta) &=Det'\left(
       \begin{array}{cc}
         -d/dt & y_i \\
         -y_i& -d/dt \end{array}
         \right)\nonumber\\
         &=Det'(d^2/dt^2+y_i^2)=\prod_{n\neq 0}(y_i^2-(2\pi
         n/\beta)^2)\nonumber\\
         &=[\prod_{n\leq 1}(\frac{2\pi
         n}{\beta}\prod_{n\leq}[1-(\frac{y_i\beta}{2\pi
         n})^2]]^2\nonumber\\
         &=(\frac{\sin\beta y_i/2}{y_i/2})^2.
    \end{align}
    When $\beta=1$, the right hand side of (3.22) is $(\frac{\sin
    y_i/2}{y_i/2})^2$.
    Therefore

    We can also use the above to evaluate
    $Det_{APBC}(\delta_{ij}d/dt+\mathcal{R}_{ij})$.
    In fact, by the above,
    \begin{equation}
    I(2\beta)=(\frac{\sin\beta y_i}{y_i/2})^2.
    \end{equation}
    This is equivalent to substitute $2n$ for $n$ in the right hand
    side of above. We also notice that the eigenvalues for
    $$
    Det'_{APBC}\left(
    \begin{array}{cc}
         -d/dt & y_i \\

         -y_i& -d/dt \end{array}
         \right)=Det'(d^2/dt^2+y_i^2)=\prod_{n\neq 0}(y_i^2-(2\pi
         n/\beta)^2)$$ are actually $n\pi\beta$ for odd $n$.

    Therefore we can have
    \begin{align}
    Det'_{APBC}\left(
       \begin{array}{cc}
         -d/dt & y_i \\
         -y_i& -d/dt \end{array}
         \right)&=Det'(d^2/dt^2+y_i^2)=I(2\beta)/I(\beta)\nonumber\\
         &=(2\cos \frac{\beta y_i}{2})^2.
    \end{align}
    Now we are in a position to get the index of the operator. We
    fix $\beta=1$ for the following calcualtions.
    $\mathfrak{D}_+$:

    \begin{align}
    ind \mathfrak{D}_+&=i^n \int_M \prod_{\mu=1}^{2n }\frac{x_0^\mu
    }{\sqrt{2\pi}}\psi_{02}^\mu\psi_1^\mu (\prod_{i=1}^n
    \frac{y_i/2}{\sin\beta y_i/2})(\prod_{i=1}^n 2\cos\frac{\beta
    y_i}{2})\nonumber\\
    &=i^n \int_M \prod_{\mu=1}^{2n} \frac{x_0^\mu
    }{\sqrt{2\pi}}\psi_{02}^\mu\psi_1^\mu (\prod_{i=1}^{2n}
    \frac{y_i}{\tan\beta y_i})\\
    &=\int_M \prod_{\mu=1}^{2n} \frac{x_0^\mu
    }{\sqrt{2\pi}}\psi_{02}^\mu\psi_1^\mu (\prod_{i=1}^{2n}
    \frac{y_i}{\tanh\beta y_i}).
    \end{align}
    We have to explain why the last two equalities hold. In fact,
    only $2n$ forms are picked up in the integrands. Thus if we multiply $i/2$
    to each entries in the curvature matrices, since each entry is a
    2-form, we then get the original $2n$ forms with an extra factor
    $i^n$.By the same reason, only when $n$ is even, the integrand
    is not zero. Therefore when $n$ is even,since $i^n=1 or -1$, the
    last two equalities hold.
    Usually we denote the integrand by $L(M)$ which is a
    characteristic class of the manifold $M$. Then eventually we
    have proved the \textbf{Hirzebruch signature theorem}

    Let's prove another famous theorem which is
    Hirzebruch-Riemann-Roch theorem. Let us review what this theorem
    says. Consider product bundles $\Omega^{0,r}\otimes V$, where
    $V$ is a holomorphic vector bundle over M,
    $$
    ...\rightarrow\Omega^{0,r-1}(M)\otimes
    V\rightarrow\Omega^{0,r}(M)\otimes V\rightarrow...
    $$
    We can apply the Atiya-Singer index theorem to this
    complex,immediately we can get
    \begin{equation}
    ind \bar{\partial_V}=\int Td(TM^+)ch(V).
    \end{equation}

    \subsection{Hirzebruch-Riemann-Roch theorem}
    Now we prove it in two steps. First, under the assumption that
    $V$ is a trivial bundle, we prove the statement by the similar
    method as Hirzebruch signature theorem. Then we use a very
    intuitive argument to show the case that $V$ is not
    trivial.\cite{AG}pointed out the method to calculate the complex when
    $V$ is trivial. We give the full calculation  here.
    Let us consider Dolbeault index on a complex manifold $M$. Here
    we have to impose a restriction on $M$ that is $M$ has a
    K\"{a}hler structure. \cite{Z} showed that on a $\sigma$-model
    defined on $M$ admits two kind of supersymmetries. The
    requirement of the K\"{a}hler structure is the sufficient and
    necessary condition to get $N=2$ supersymmetry.

    We can decompose the exterior
    algebra by the direct sum of holomorphic $p,q$ forms by
    \begin{equation}
    \Omega^*(M)^\mathcal{C}=\bigoplus_{p,q=0}^n \Lambda ^{p,q}(M),
    \end{equation}
    where $\Lambda^{p,q}$ denotes forms with $p$ holomorphic indices
    and $q$ antiholomorphic indices.

    We can get the index of nontrivial $V$ by the following
    intuitive argument. First of all, since the index is an integer,
    the index remains the same when the spin field or the gauge
    field underlies a continuous deformation. In other word, mathematically, the
    index only depends on the fiber bundles of the spin field and
    the gauge field up to isomorphisms of fiber bundles. Therefore
    the index is equal to the integral of some index density which
    only depends on a local characteristic class that involves the
    two fiber bundles.We are able to find local coordinates such
    that the nontrivial charts of the two bundles don't intersect.
    In the local charts in which spin bundle is nontrivial, the
    index density is identical with the case that there is no gauge
    field at all \cite{AG}; in the local charts in which the gauge field is
    nontrivial, the index density is just $ch(V)$ \cite{F1},\cite{F2}. In the
    local charts in which both fields are trivial, the index density
    is exactly 1.There is no conflict. Therefore we retrieve the
    Hirzebruch-Riemann-Roch theorem in a complex manifold with a K\"{a}hler
    structure.

\section{conclusion and further questions}
  The idea of calculating using supersymmetry is kind of simple although the real calculation is complicated.
  The most important thing in calculating the index is to clearly
 indetify the supercharge for the operator and find the corresponding physical
 theory. In the process of proceeding section, it seems hard to calculate the index of a Dirac
 operator coupled twisted by a gauge field from a supersymmetric
 theory. It is easy to deal with using Fujikawa's method. The reason
 that the index of a Dirac operator in a spin field is easy to deal
 with in a supersymmetric theory is that supersymmetry actually
 exchanges the fermionic varialbes and the bosonic
 coordinates. Therefore supersymmetry founds a connection between
 the Dirac operator and the geometry of the manifold which is just
 the spin field. The difficulty lies that supersymmetry doesn't
 provide the connection between a gauge field and the geometry of
 the manifold directly.

In addition, there is still a question. In the proof of the
Hirzebruch-Riemann-Roch theorem, we assumed that the complex
manifold has a  K\"{a}hler structure, but the original theorem
 doesn't require the existence of a K\"{a}hler structure. Since the
 K\"{a}hler condition is a necessary and sufficient of condition to
 yield supersymmetric $\sigma$-model for a manifold, a question
 arises which is how to prove the original theorem which is without
 the K\"{a}hler condition using supersymmetry.


\end{document}